# Web API Fragility: How Robust is Your Web API Client

Tiago Espinha, Andy Zaidman and Hans-Gerhard Gross

**TU**Delft

SERG

# Web API Fragility:
# How Robust Is Your Web API Client?

Tiago Espinha
Delft University of Technology
The Netherlands
t.a.espinha@tudelft.nl

Andy Zaidman
Delft University of Technology
The Netherlands
a.e.zaidman@tudelft.nl

Hans-Gerhard Gross
Esslingen University
Germany
Hans-Gerhard.Gross@hs-esslingen.de

*Abstract*—Web APIs provide a systematic and extensible approach for application-to-application interaction. A large number of mobile applications makes use of web APIs to integrate services into apps. Each Web API's evolution pace is determined by their respective developer and mobile application developers are forced to accompany the API providers in their software evolution tasks. In this paper we investigate whether mobile application developers understand and how they deal with the added distress of web APIs evolving. In particular, we studied how robust 48 high profile mobile applications are when dealing with mutated web API responses. Additionally, we interviewed three mobile application developers to better understand their choices and trade-offs regarding web API integration.

## I. INTRODUCTION

Modern-day software development is inseparable from the use of Application Programming Interfaces (APIs) [1]. Software developers access APIs as interfaces for code libraries, frameworks or sources of data, to free themselves from low-level programming tasks and/or speed up development [2]. In contrast to statically linked APIs, a new breed of APIs, so called web service APIs, offer a systematic and extensible approach to integrate services into (existing) applications [3], [4]. However, what happens when these web APIs start to evolve? Lehman and Belady emphasize the importance of evolution for software to stay successful [5], and updating software to the latest version of its components, accessed through APIs [6]. In the context of statically linked APIs, Dig and Johnson state that *breaking changes* to interfaces can be numerous [6], and Laitinen says that, unless there is a high return-on-investment, developers will not migrate to a newer version [7].

When integrating with a web API however, developers can no longer afford the inertia that was noted by Laitinen. The web API provider sets the pace for migrating to newer versions (eventually removing older ones altogether) and client developers are forced to migrate. In the statically linked API context, developers could choose to stay with an older version of e.g. libxml, which meets their needs, yet, with web service APIs the provider can at any time unplug a specific version (and functionality), thus forcing an upgrade.

Indeed, while Laitinen claims that client developers will postpone migration to newer versions until there is a high return-on-investment, in previous work [8] we found some web API providers are eager to push breaking changes and force client developers to migrate to a newer version within a period as short as 4 months.

Through their loose coupling [9] and REST interfaces, web APIs can easily be integrated into applications with a single HTTP request [10]. However, as the integration becomes as simple as exchanging HTTP requests, do client-side developers consider the consequences of ever-evolving web APIs?

We choose to perform our investigation into web APIs in the realm of mobile applications. This was a conscious decision instigated on the one hand due to the linear growth of mobile Internet usage[1], and on the other hand because mobile apps connected through web APIs are an integral part of the mobile computing experience [11], thus ensuring relevance.

Being aware of the ever-growing importance of and reliance on web APIs [12], particularly in the mobile computing domain, we wonder how well-prepared some of the most popular Android mobile applications are with regard to a number of factors which include:

- changes in the web API response from the server due to evolution of the web API
- interrupted HTTP requests due to e.g. loss of Internet connectivity
- empty response messages due to server overload

In order to steer our research, our main research question is *how well-prepared are Android mobile applications with regard to changes in response messages from the web API* which is then divided into the following sub-research questions:

[RQ1] How robust are mobile apps when the web APIs being used return unexpected responses?

  [RQ1.1] How can we simulate unexpected responses from web APIs?

[RQ2] Have web API client developers developed resilience against changes in the web API or failure of the web API?.

To address these questions we performed a study on 48 Android mobile applications which make use of at least one web API. In our study we perform a series of mutations aimed at mimicking potential real-world scenarios where the web API changed its behavior either through (communication) failure or software evolution and observe how the different mobile applications fare in dealing with such mutations.

---

[1]Already in the United States of America, reports show that specifically mobile applications' Internet usage has in 2014 surpassed Internet usage on desktop computers — http://money.cnn.com/2014/02/28/technology/mobile/mobile-apps-internet/, last visited June 6th, 2014.

The remainder of this paper is structured as follows: in Section II we describe our approach, namely the details of using mutation analysis for analyzing web API client robustness, in Section III we describe our experimental setup including how the mobile applications were selected, the added dimensions we analyzed and details regarding the developer interviews, Section IV describes the results of our experiment with mutation analysis as well as the insights from the developer interviews, Section V discusses potential threats to validity of our work and lastly we present our conclusions and future work in Section VII.

## II. Approach

In order to investigate the robustness of Android mobile applications with regards to possibly fragile web APIs, we first start by explaining our approach to simulate unexpected response messages from the web API, which is loosely based on mutation analysis, also found in the area of software testing [13]. In this section, we first introduce our mutation analysis in Section II-A, after which we explain the technical setup that we have used to apply the mutations in Section II-B.

### A. Mutation Analysis — What to Mutate?

Our mutation analysis consists of changing the web API response for a particular web API request sent by a mobile application. While previous research by Bozkurt et al. [14] touches upon the issue of XML and SOAP perturbations, their approach and goal are focused on identifying faults on the *web service/API*. While our research also makes use of message perturbation, we apply such perturbations (referred in this paper as mutations) in the *response* received from the web API as to test the robustness of the mobile application which initiated the conversation. To do so, we first start by selecting a repeatable action within the application and capture a valid response for that action using Charles [2], a web debugging proxy. This response is then used as a basis for the mutations which are also applied using Charles' rewrite feature.

Xu et al. [15] set forward four perturbation primitive operators for XML documents: two insertion operators and two deletion operators where the difference is the position in the XML tree where new nodes are added and deleted. One of the addition operators which inserts nodes at the same level as existing nodes is also included in our study. The other addition operator relates to datatype insertion and since none of the studied web API responses contain datatype definitions, was excluded from our study. The same reasoning was applied to the deletion operators.

Thus, we devise two operators from the aforementioned work: addition of new unrelated nodes and deletion of existing nodes (referred to in this paper as *field addition* and *field removal*, respectively). We extend these operators with four other operators: *malforming a response*, *replying with an empty message*, *changing the implicit data type of a field* and *disrupting the data formatting*.

**Malformed responses** can happen for a number of reasons. As an example, if the data encoder of the web API fails to properly sanitize strings, it could lead to floating reserved characters which in turn break the document's data format. This can happen while, for example, encoding a JSON string which contains double quotes and these are not properly escaped (i.e. "foo":"b"ar" as opposed to the valid "foo":"b\"ar"). In order to mimic these failures, our malformed response mutation consists of mutating the web API response as to make it malformed in its respective data format. In XML we mutate the response by breaking an XML tag (e.g. '<data>' becomes '<data') and in JSON this is achieved by leaving a dangling double-quote in a JSON string (e.g. "foo":"bar" becomes "foo":"bar).

**Empty responses** can be a symptom of different types of ailments on the web server. For instance, should the web server be at the edge of its maximum capacity, some requests may receive an empty response. Similarly, if the connection is terminated due to communication issues, it could also lead to empty responses being returned to the mobile application. Our empty response mutation consists simply of replacing the response with an empty-bodied HTTP response.

Whether using a structured approach such as semantic versioning [3] where only major versions are allowed to bear breaking changes, or using a more lenient approach, when dealing with a web API it is possible that some fields are removed. Indeed, even renaming a field will appear as an addition plus a removal and would constitute a breaking change. Breaking changes caused by the removal of fields can be found in our previous work [8], in particular regarding the Facebook web API [4]. Also Li et al. [16] show that indeed more providers rename parameters which also results in breaking changes. In the context of our mutations, a **field removal** mutation consists of removing one or more fields from the web API response. In most cases only one field is removed and, should the application still function normally, we take it one step further by removing more fields until it crashes or no more data is available to be removed. Of note is the fact that despite removing fields, special care was taken to maintain the web API response semantically valid (as opposed to the aforementioned malformed response).

Also as a part of software evolution, web APIs may at some point change the (implicit) data types of certain fields. For instance, while the price field can be a string which includes the currency symbol, at a later stage the price field can become a purely numerical field. To address such cases, our mutation of **changing data types** consists of two changes: replacing an integer with a string and whenever possible, the reverse operation. As all the mobile applications run on Java (by force of the Android platform) and since Java is a statically typed language, if special care is not had when parsing the web API

---

[2]Charles Web Proxy — http://www.charlesproxy.com/

[3]Semantic Versioning — http://semver.org/

[4]Facebook Completed Changes — http://bit.ly/fb-completedchanges

response, type mismatches could occur.

Lastly, the **field addition** mutation consists of adding one or more fields to the web API response which contain irrelevant data for the mobile application. The **data formatting** disruption consists of adding line breaks and white space as an attempt to verify whether the mobile applications are sensitive to this type of changes.

*B. Mutation Analysis — How to Mutate?*

In order to mutate the responses from each mobile application's respective web API, we start by installing the chosen applications for our study on a Google Nexus 7 tablet (running Android KitKat 4.4.2) and configure the tablet to redirect all the network traffic through a transparent proxy (Charles Web Proxy) setup in a separate machine on the same network.

On a first instance, while using the transparent proxy, we identify a reproducible action which causes a request and response interaction with the web API. We then collect a standard response (all the standard responses used are available online [5]) for that particular web API request made by the application under study and configure the transparent proxy to replace the response of all other similar requests (for the purposes of the Charles web proxy, similar requests means requests sent to the same endpoint) with a customized response. The customized response is in fact the original web API response although slightly modified. We modify the response as to verify that the application executes as expected and that the new data is being loaded into the mobile application.

The original web API response is then further disturbed with a number of different types of mutations (explained in detail in section II-A). For each mutation we observe how the Android application reacts to such changes. Our observations are then categorized into different types of behaviors (e.g. crashing or indefinitely loading without a timeout) and turned into a report (a sample is available online [6]) which is two-fold in its content: it starts with a statistical overview displaying how many applications behave in each of the identified categories, and culminates with a report specific to that particular application.

## III. EXPERIMENTAL SETUP

For our experiment, we first need a body of mobile applications to be analyzed. How these applications were chosen is described in detail on section III-A. We then introduce mutations in web API responses through the use of a transparent web proxy, described in subsection II-B. These mutations, described in detail in subsection II-A, range from slight deviations from what would otherwise be a valid response all the way to making the responses malformed as an attempt to emulate breaking changes and faults on the web API.

We also expand this quantitative view with a more qualitative approach. To do so we interviewed 3 developers, each from a different application under study. More details regarding these interviews are also in the subsections below.

---

[5]Web API Responses — http://bit.ly/figsharescam2014
[6]Sample status report — http://bit.ly/report-web-api

*A. Application Selection*

With our study we aim to understand what is the current state of web API integration issues. We do so by analyzing some of the most popular and widely used Android applications. All the applications under study were required to meet two criteria.

The first criterion is that each application must use at least one web API. Important to note is that our definition of web API excludes mobile applications which simply load HTML or RSS feeds. These exclusions are due to the fact that with an HTML response, no processing is required on the mobile application. With this we exclude also any mobile application which may use screen-scraping techniques [17]. Similarly, RSS feeds have a fixed structure which means they are not susceptible to software evolution changes which make them not applicable for our study.

The second criterion is related to how the mutation analysis is applied in the web API response. Tampering with such web API responses is only possible if the web API communication happens over an insecure channel such as HTTP. Indeed, this is particularly important as an encrypted protocol such as HTTPS does not allow for changes to be made to the content of its messages (as this is indeed its purpose).

For this study we picked candidate projects from the top 100 free applications available in the Google Play Store of the Netherlands, United Kingdom, United States of America, Canada, Australia, Belgium, Brazil, Spain, Germany and France. In total we installed 198 projects of which not all were usable. For instance, 65 use HTTPS and a number of other projects use proprietary binary data formats for the communication. Similarly, other projects which use compressed files as a means to transfer the web API responses (e.g. ZIP files) were not mutable using our approach as it would require the response to be decompressed and recompressed on the fly, which is not possible using Charles Web Proxy. Ultimately, we were able to compile a list of 48 projects (Table I [7]) which were all analyzed and compose the corpus of our study.

Of note is the fact that all of these projects make use of either XML or JSON as a data format for the web API requests. This is particularly relevant as a data format like XML requires an additional XSD schema document to enforce data types. For this to happen, the XML document would require references to the XSD [8] document which can be used to validate it. Such cases have not been encountered in this study which means no XML documents were being (explicitly) validated against an XSD schema. During our mutation analysis, changing data types was still attempted on XML documents (i.e. changing a field with a numeric value to a string) even though none of the fields were specifically numeric as it happens with JSON where numeric fields can be identified by the lack of quotation marks.

---

[7]The Buienalarm and Eurosport mobile applications make use of two distinct web APIs and therefore appear twice on the list.
[8]XSD Schemas — http://www.w3.org/TR/xmlschema11-1/

## B. Caching and Versioning

After beginning our experimental study, we found that for some applications, after the *"sample web API response"* was collected to be used as basis for the mutations, the mutated data (even with a string for string mutation) would not be loaded. This hinted at the usage of caching on some of the mobile applications. As caching is of particular interest for mobile applications where the Internet connectivity may sometimes experience slow bandwidth and where the web API may not respond due to high peaks of server load, we collected data on whether each mobile application uses caching and whenever possible, how is the caching used.

On what concerns versioning, in previous work [8] we found that some of the high-profile state of the practice web APIs (e.g. Twitter, Google Maps, Netflix) make use of some form of versioning. Still, we also found a major web API provider (Facebook) which does not make use of any form of versioning in their web API. Having studied these two different approaches and how client developers perceive each of the aforementioned web APIs, we also collect data on which web APIs are versioned and which type of versioning is used.

## C. Developer Interviews

While the empirical study described above provides interesting insight on how a large body of mobile applications react when web APIs experience different failures and changes, it does not provide an explanation as to the choices of their respective developers. As an attempt to shed some light on the developer perspective of developing and testing an application which integrates with a web API, we aimed at interviewing the developers of some of the mobile applications under study. These interviews took the format of a semi-structured interview [18] using the questions in Table II as a basis to stimulate the exploratory discussion.

We selected 14 applications which stood out either due to a special versioning mechanism, because they crashed or because of some particular behavior that other mobile applications did not demonstrate. For these select applications we composed an application-specific *status report* with our findings. These reports were then sent to their respective developers along with an invitation to participate in an interview.

Ultimately we were able to interview three software developers: the mobile software architect for OZsale, the product manager for the *Trivago* mobile application and lastly, the third developer is the sole developer of the *NS.nl* Android application. The interviews lasted 15 minutes on average.

The *status report* we compiled and sent to the developers contains statistical information on all the projects under study. We provide data on how many projects crash due to a mutation, how many use versioning and the divide between JSON and XML implementations. Also for each type of mutation we provide statistics on the different observed behaviors. We also present statistics on how many applications use caching and complement the report with application-specific findings (e.g. some applications still load malformed data). It is also these outlier findings of behaviors that are not common which we aim at clarifying with the developers through the interviews.

| Application | Versioning | Caching |
|---|---|---|
| NS.nl | ✓ | ✓ |
| yr.no | | |
| Skyscanner | ✓ | |
| Buienalarm (own API) | | |
| Buienalarm (OpenWeatherMap) | | |
| BBC News | | |
| IMDB | | |
| Daily Mail Online | ✓ | |
| WeatherBug | ✓ | |
| TheWeatherChannel | | |
| Reddit is fun | | |
| Ozsale | ✓ | |
| Wish | | |
| tramTracker | | |
| NRL - League Live | | |
| The Masters Gold | | |
| mobile.de | | |
| Onefootball | ✓ | |
| Wetter App | | |
| MeinProspekt | ✓ | |
| Wetter.de | ✓ | |
| TV Movie | ✓ | |
| Wetter.com | | |
| McDonald's Deutschland | | ✓ |
| H&M | | |
| eltiempo.es | | |
| Liga de Futbol Professional | | |
| TecMundo | ✓ | |
| Trivago | ✓ | |
| Resultados Futbol | | |
| La Chaine Meteo | ✓ | ✓ |
| Resultats Foot en Direct | | |
| RATP | | |
| ViaMichelin | ✓ | |
| Le Figaro | ✓ | ✓ |
| Le Parisien | | ✓ |
| Eurosport (XML) | | |
| Eurosport (JSON) | | ✓ |
| Tele Loisirs | | |
| NU.nl (stocks) | ✓ | ✓ |
| Just Eat | | |
| Questionmark | ✓ | |
| Couverts | | |
| Trulia | ✓ | ✓ |
| 24Kitchen | | |
| NL Treinen | | ✓ |
| Huizen | ✓ | ✓ |
| Kieskeurig | ✓ | |
| Jumbo FoodMarket | ✓ | ✓ |
| Pull&Bear | ✓ | |

TABLE I
LIST OF MOBILE APPLICATIONS STUDIED

| | |
|---|---|
| Q1 | What was the design decision behind choosing HTTP over HTTPS? |
| Q2 | Why are no caching mechanisms used? |
| Q3 | Is versioning not used for a particular reason? |
| Q4 | Is the web API used by other (third-party and or mobile) applications? |
| Q5 | Is the mobile application native to Android or generated with a mobile development framework? |
| Q6 | Is the mobile application developed by the same team as the web API? |
| Q6.1 | In particular when it is not developed by the same team, how does the mobile application team learn about the web API changes? |
| Q7 | How frequently are the web API and mobile application updated? |
| Q8 | Have there been problems in the past with breaking changes causing the mobile application to break? |
| Q9 | Are there automated tests in either the mobile application or web API (unit tests, integration tests, etc)? |

TABLE II
QUESTIONS ASKED DURING THE DEVELOPER INTERVIEWS

## IV. EXPERIMENT RESULTS

In this section we present our findings regarding each observed behavior in the Android applications under study upon applying the different mutations to the web API response. Specifically, we report on four of the six initially proposed mutations. Two of the mutations, the field addition and data formatting, represented no problem for any of the studied applications and are thus not further discussed in the results below. We provide an analysis of the different behaviors displayed by the mobile applications under study and whenever relevant, provide anecdotal examples.

We also expand upon the results of the mutation analysis with data on the different types of data caching and web API versioning encountered as well as developer input on some of the choices used in web API integration. Of note is that the results are valid for the respective versions studied, current as of June 2014. As each of the mobile applications and their respective web API may change, so may the mobile applications' reaction to web API mutations.

### A. Reactions

*1) Force Close:* On the Android platform, uncaught exceptions will cause the application to crash and be immediately closed (force closed). This is always caused by a Java runtime exception, thrown in the context of that specific application, which was not correctly handled.

In our study, each of the applied mutations caused at least one application to force close. Table III shows an overview of how many applications force close according to each mutation.

**Malformed Response**. The only application which demonstrated this behavior (*Wetter.com*) will immediately crash as soon as the malformed data is required in the user interface. One other application (*Le Parisien*), despite not force closing, would show a message upon loading the data that would force the application to exit itself thus rendering the application unusable. The reduced number of applications force closing implies that for the majority of the applications, some validity checking is being done to the web API response.

**Empty Response**. As a response which may happen in the event of a server overload, it is surprising that not all applications were able to deal with an empty message. Indeed, the *TecMundo* application immediately crashes upon receiving an empty response. In fact, when such a mutation was applied, the application would become unusable by crashing on startup.

**Field Removal**. While the mobile applications were in general resilient to malformed and empty responses, removing fields from the web API response caused 10 applications to force close. An interesting example is that of the *NS.nl* mobile application. While the web API response does not specify optional arguments, the removal of some arguments still allows the application to load the response normally without crashing or showing an error message. However, as further (potentially non-optional) arguments were removed, the application would crash every time we attempted to load the incomplete data. Of note is the wide user-base of this particular application which, according to the Google Play Store statistics, has been installed between 1 million and 5 million times.

In our study we have also found another application which makes use of the same web API as *NS.nl*, the *NL Treinen 2 - NS*. An interesting finding regarding this third party application is that when suffering the exact same mutations the official *NS.nl* application had endured, it never crashed. This added care when dealing with the web API may in fact be due to the web API being provided by a third party (from the *NL Treinen 2 - NS developers'* point of view) as opposed to what happens with the official NS.nl mobile application. Indeed, upon interviewing the developer for the *NS.nl* application, we learned that the interviewee is the same developer who develops the web API. Unfortunately, the developer for *NL Treinen 2* did not react to our interview request and we can thus not fully confirm our hypothesis.

**Changing Data Type**. Also a cause of force closes is changing the data type of some fields in the web API response. In particular, this caused 3 applications to crash. While we were not able to investigate the exception being thrown, it is possibly related to the parsing of the message and of a type mismatch between Java's statically typed variables and the strings being parsed into different types.

**In sum**. With the exception of the *field removal* mutation, less than 10% of the mobile applications crashed when facing the different mutations. More worrying is the 10 applications which force close upon the removal of fields from the web API response. Indeed, field removal is a potential change which evolving software faces through e.g. refactoring of the web API response. Our results reveal that potentially there are no negative tests scenarios being applied or the existing tests do not cover scenarios such as a malformed response.

*2) Error Message or Silent Failure:* Showing an error message is a graceful way of letting the end-user know that something did not go as expected and what his or her course of action should be (e.g. try again or check the Internet connectivity). This behavior is, in all cases, preferred to letting the application crash and close itself.

In our study we do not make a distinction as to which approach may be better at notifying the end-user. Nonetheless, the content of the error messages should be detailed in explaining the cause of the issue and what should the end-user do about it, indeed, in the dynamic domain of web APIs where new versions are released regularly and often without the knowledge of the client developers, versioning mechanisms could help in delivering better advice to the end-user (e.g. update the mobile application to the latest version or simply retry). Our findings show that whenever the error message exists at all, these are vague and do not offer insight about what

| Mutation | # of apps |
|---|---|
| Malformed Response | 2 applications |
| Empty Response | 1 application |
| Field Removal | 10 applications |
| Changing Data Type | 3 applications |

TABLE III
MUTATIONS VERSUS FORCE CLOSE

went wrong. Anecdotal evidence of this is provided below.

Also, in our study, not all the mobile applications make use of alerting mechanisms to inform of issues with the web API response. In fact, we contrast the showing of an error message with a silent failure. By not showing any reaction to the end-user's input, the mobile application may induce confusion. More so than by showing a generic error message which will at least let the end-user know something went wrong. Indeed, when the application simply does not react, it is impossible for the end-user to know whether the data is still being loaded (as in these cases there was also no visual indication of loading) or whether he or she should try again to refresh the data. Table IV shows an overview of how many applications show an error message or silently fail according to each mutation.

**Malformed Response**. An example of one of the native error message mechanisms is shown by the *NS.nl* Android application which reports that it *"cannot retrieve data from server"*. This application's error message, while better than remaining mute about the failure (as can be seen in the next paragraph), also serves as an example of a generic message which offers no indication of how the end-user should proceed. In fact, the data was retrieved from the server but an error happened while processing it.

In contrast, some mobile applications under study did not react upon facing a malformed response. Also in this type of behavior, two different types arise. In some applications such as *IMDB* and *Daily Mail Online* a *"results screen"* is shown with zero results, which does at least hint that the loading has stopped (even though no results are shown while the web API response did contain results). In contrast, the *Resultats Foot en Direct* application does not have a dedicated results screen and thus simply a void is shown where the results should appear. This will either leave the end-user waiting longer than necessary or trying to repeat the action multiple times.

**Empty Response**. When dealing with an empty response, 16 applications of those under study did show an error message. Of special note are 4 of these applications (*mobile.de*, *Resultados Futbol*, *Couverts* and *Trulia*) which show a message claiming that no results were found and that the end-user should change his or her search criteria. This reveals that the application did not recognize the empty response as a fault, even though an empty set of results from the web API was in neither of the applications an empty message (i.e. even when no results were available, some boilerplate JSON or XML encoding still should be returned). The remaining majority of the 16 applications showed generic *"network error"* messages, with special attention to the *yr.no* application. This was the only one which actually reported an *"empty response"*. Still, the most common behavior seen amongst the 48 mobile applications and one that affects more than two thirds (28 out of 48) of the applications under study is a silent failure. More specifically, these mobile applications would stop loading and never present the user with a message describing why the loading had stopped and why no data was loaded.

| Mutation | # of apps ⇒ error message | # of apps ⇒ silently fail |
|---|---|---|
| Malformed Response | 14 applications | 34 applications |
| Empty Response | 11 applications | 37 applications |
| Field Removal | 6 applications | 42 applications |
| Changing Data Type | 5 applications | 43 applications |

TABLE IV
MUTATIONS VERSUS ERROR MESSAGES

| Mutation | # of apps w/timeout | # of apps indefinitely loading |
|---|---|---|
| Malformed Response | 40 applications | 8 applications |
| Empty Response | 39 applications | 9 applications |

TABLE V
TIMEOUT

**Field Removal**. Similar to what happens with an empty response, 6 out of the 48 mobile applications under study silently fail whenever one or more fields of the response message are removed. A silent fail means that the application does not show any of the remaining data still available in the response, does not crash and does not show an error message indicating what went wrong with the request.

The remaining applications which *do* show error messages, just like in the malformed response mutation, show no specially tailored error message to cover the lack of certain data in the web API response.

**Changing Data Type**. As with the other faults, some of the mobile applications silently fail without informing the end-user about the fault. In the case where data types were changed, 5 out of the 48 applications did not report an error even though the data would not load. The error messages that did get shown were generic and did not pinpoint the problem.

**In sum**, a considerable number of mobile applications under study do not make use of error messages to inform the end-user of problems with the web API. Indeed, some of the mobile applications simply silently fail and others react as if no *"search results"* were found. Furthermore, with exceptions such as the *yr.no* application which clearly points out which was the fault at hand, most applications show the same generic error message for all the different mutated web API responses.

*3) Timeout:* Timing out is an important part of reporting a failure. Indeed, when presented with malformed and empty responses, some of the applications under study (Table V) do not time out. This means that these applications will stay indefinitely *"loading"* (although nothing is really being loaded) until the application is closed or the end-user touches the screen. For the applications under study, only malformed and empty responses caused applications to not time out. In this event, the application does not force close (thus the end-user does not know that something went wrong) nor does it show an error message. It is then up to the end-user to decide when to stop waiting and close the application.

**Malformed Response**. When dealing with malformed responses, 8 applications out of those under study do never time out. In this case, the mobile application is left stuck in a loading screen. While the absence of a force close indicates that whatever exceptions which arise from processing a malformed response are being handled (since the application did not crash), the handling of the situation altogether is not ideal as the applications never time out. If it is the case that the document parser does not report whenever the documents are damaged, then a different parser would be preferable. While we cannot confirm either situation on the source code, it is either handling and muffling the exception or it does not know altogether that the message is damaged. In either case it never closes the loading screen and fails to provide the end-user with insight as to what happened.

**Empty Response**. One of the most common behaviors experienced when replacing the web API response with an empty message was the application hanging indefinitely, never timing out or informing the end-user that the loading had failed. While it is not possible to verify the reason for this behavior in the applications' source code due to their closed source nature, it is likely the affected applications always expect a reply with content. When the content is not present, the applications hang until there is content (which would then never happen). In fact, RESTful web APIs may indeed at times reply with an empty message, for example with HTTP status codes of 304 (Not Modified) or 204 (No Content) [9]. In our study we experienced web APIs which replied with an empty HTTP message having a status code of 301 (moved permanently) which indicates that a request should be made to a different URL. In our study, 17 mobile applications out of the 48 analyzed presented this behavior and would not stop loading until the screen was touched or a key was pressed.

**In sum**, when dealing with a web API where a network connection is involved, timeouts are of utmost importance. In our study we found a number of applications which never time out and indefinitely leave the end-user thinking something is still being loaded when in fact the application simply did not handle the mutated web API response correctly.

### B. Data Caching

While applying the mutations to the web API responses, some applications would not initially attempt to load the mutated data. This was due to the fact that because the data had just successfully been loaded (from our first execution, probing for a testable action), the data would be stored in cache. Until the cache timeout was met, new data would then not be loaded. In some applications, even though caching was used, a request would still be sent to the web API but the response would be discarded. While this approach to caching does not help with minimizing web API interaction and thus network data usage, it may help as a backup whenever network connectivity is not available.

The finding that some applications were indeed using caching led us to investigate how many of the mobile applications under study were making use of it. Out of those that did use caching, we tried to find out how was it performed. The use of caching presents a trade-off between different factors: while the end-user does not always have the most recent data at his or her disposal, with caching the number of requests being sent to the web API can be reduced. Also in the cases where connectivity is temporarily lost, which is a potential risk when using wireless networks, having a cache means that the end-user can still access *some* data and have some degree of interactivity with the application. With this in mind, caching data does help to some degree in dealing with a web API whose responses should not be taken for granted (as opposed to what happens with a statically linked API).

Our results show that the majority (34 out of 48 applications) do not use caching. Amongst those that do, one application in particular stands out (*McDonalds Deutschland*). While from an external perspective we assume most of the applications perform time-based caching where the cached result times out after a set amount of time, the *McDonalds Deutschland* mobile application implements caching by hashing. Before invoking the web API method it requires, it first queries the web API for a checksum hash of the web API result. The hash, which is significantly smaller than the actual response, is compared to the hash of the previously queried data and if it matches, no new request is made. This type of caching helps reduce mobile data usage and it improves the response time of the mobile application.

For all the other applications which make use of caching, quitting the application would not help in clearing the cache. In fact, we had to manually clear all the data of the Android application using the built-in Android data clearing function which resets the application to the state as if it had just been installed and never executed. We therefore assume these applications implemented time-based caching where after a fixed period of time the cached data would be marked dirty and a new request could be issued to the web API.

In contrast, the cache for one other mobile application studied (*Huizen*) times out as soon as the application is closed.

### C. Versioning

An important aspect when dealing with libraries in general is the usage of versioning. It is widely used by build tools like Maven where a version of each library must be specified.

When it comes to web APIs, however, many web APIs still do not make use of versioning. Indeed, our study shows that only 16 out of 48 applications' web APIs (less than half) make use of some form of versioning.

In previous work [8] we highlighted the importance of versions in the web API context. Especially when the client developers have no control over when changes happen to the web API behavior, versioning that behavior allows the client developers to know what behavior to expect from a particular web API. The most widely used versioning mechanism observed in web APIs used by the mobile applications under study is of specifying the version in the URL (e.g. www.weather.com/v1/report). This way, whenever migrating a client's integration from one version of the web API to another (either as an attempt to use features provided in the new version or merely pushed to do so by the web API provider), the client developer knows some changes are to be expected. Some applications take this practice further and make use of semantic versioning. One of the applications under study (*OZsale*) does use a variant of semantic versioning. Currently at version 3.4, minor version increments (e.g. all the different versions within 3.X release) should not in principle introduce breaking changes and are thus safe to interchange.

---

[9]REST Patterns, HTTP Status Codes — http://bit.ly/restpatterns

What is then a surprise is how such a high percentage of mobile applications make use of the web API without any form of versioning. Ultimately when such a web API introduces changes, all the clients which have not yet migrated to the latest version will still be interacting with a changed web API, which may not be compatible. Evidence of a scenario where this would potentially happen would it not be for versioning comes from one of the interviewed developers. The developer interviewed in the context of the *OZsale* application claimed that indeed, some end-users do not update their mobile applications and that 5% of their user-base (of 100.000∼500.000 users according to the Google Play Store) was still using their mobile application's very first version.

### D. Developer Interviews

In this section we present the findings gathered from interviewing the three participants in our study. In the paragraphs below we refer to the questions shown in Table II.

**Insecure HTTP**. Referring to question **Q1** regarding the design decision of using HTTP over HTTPS, an intriguing finding of our study is how such a large number of mobile applications (indeed, all the 48 under study) still make use of insecure HTTP. Data sent over HTTP allows for the data to be both eavesdropped upon and, indeed, tampered with in the same fashion as is performed in this study. When confronted with this question, one of the developers claimed he did not in fact know why their web API was using HTTP as the application protocol because the web API is developed by a different team. However, he mentioned the use of HTTP was likely related to the Chief Technology Officer being *"obsessed about performance"* where HTTPS does indeed show its downside with an overhead on both processing (as each request is encrypted server-side) and on network overhead (as the encrypted request grows larger and has a lower compression ratio of encrypted data). Furthermore the same developer claimed that despite using HTTP for interactions which are not sensitive from a security standpoint, their mobile application does make use of HTTPS for login and payment interactions.

**Caching**. While all the interviewed developers perceive caching (question **Q2**) as a useful mechanism to reduce network usage, specifically the developer of the *NS.nl* application raised a concern about the necessity for *"fresh data"*. Indeed, this application provides information on the Dutch train departure and arrival times which are at times susceptible to delays. It is thus crucial to always display the latest data.

Another one of the interviewed developers (the mobile software architect for *OZSale*) justified the lack of caching as it being a lower priority requirement. While such a feature is already present in the iOS version of the mobile application, at the time the Android version started being developed *"the libraries available for caching in the Android platform were not yet mature enough"*. The iOS application goes a step further and makes all of the data available for offline browsing.

Also the developer responsible for the web API at *Trivago* claimed that caching would in fact stay in the way of the mobile application's performance for their specific case. Caching would require the mobile application to keep track of which data it has available and only request the delta between what it already has and the results it still needs to fetch. To do so with caching and without state would make for chatty communications. The mobile application would have, with every request, to report what it already has in cache and what it requires. *Trivago* contains a more pragmatic approach where sessions (i.e. stateful exchanges) are used which allows the cache to be on the server-side and thus lower the chattiness which is desirable for both performance and data usage.

**Versioning**. Two of the three interviewed developers (for the *Trivago* and *OZsale* mobile applications) have versioning mechanisms implemented in their respective web APIs.

For instance, the *Trivago* web API makes use of HATEOAS [10] (Hypermedia as the Engine of Application State) versioning approach. The HATEOAS approach makes use of HTTP headers (Accept-Type and Content-Type) as a way to handle versioning and description of the data and since it stands central to being *the* way a RESTful web API should be versioned, Liskin et al. [19] have in fact devised an approach to *"add HATEOAS support [to services] as an afterthought"*. When asked about why this particular versioning mechanism was used, the answer was that even though the Trivago web API is still in its first version, HATEOAS was purposefully chosen as a way to future-proof the evolution of the web API.

The developer of the *OZsale* application also stressed the importance of their versioning system. While the data itself is not versioned (as it happens with HATEOAS), a version number must be used in the URL to inform the server of which version of the web API the mobile application requires.

An interesting divide between the two aforementioned developers is how old versions of the web API are handled. The Trivago software architect underlined the fact that it is costly to maintain different versions in parallel - which they try to avoid at all costs, while the *OZsale* developer claimed that the different versions were a core part of their different platforms. More precisely, while the website was running on the latest version of the web API, the different mobile platforms were lagging at least 5 minor versions behind (all of which were still available and fully functional). The reasoning for this was the existing delay between submitting a new version of the mobile application to the respective application store and the application actually being available (e.g. the developer claimed that in the iOS App Store this delay can be up to 1 week).

The developer of the *NS.nl* mobile application claims that it was not necessary to have a versioning mechanism in place for because not many changes are pushed to the web API. In particular, over the course of four years of development, no breaking changes have been applied to the web API.

**Evolution & Communication fragility**. Another interesting finding is anecdotal evidence of communication issues between the different teams involved in the development process. Indeed, one of the developers claimed that at least twice in their project, changes were pushed to the web API which inadvertently broke backwards compatibility. The result was having a mobile application which was crashing. This anecdote raises an issue which is also supported by our analysis of the 48 mobile applications: not all mobile applications are built with the consideration that the web API can change *at any time*. This was especially relevant as in this very same project, changes were being pushed daily with breaking changes taking place every two months highlighting the need for excellent communication between the mobile teams and the web API team should these teams not be one and the same (questions **Q4**, **Q6** and **Q6.1**).

---

[10]Versioning REST Services — http://bit.ly/versioningrestservices

**Integration Testing**. While using a static library it is possible to test it and expect it to behave the same. However, when using web APIs where the behavior can change due to a simple patch which fixes what was buggy (but expected) behavior can cause the mobile application to suddenly misbehave. This highlights the importance of both positive and negative testing, that is testing both scenarios which are part of the use cases as well as unexpected but potential scenarios.

Our empirical data suggests that Çalıklı and Bener's [20] observation on confirmation bias regarding testing may indeed affect some of the studied applications. Indeed, while some of the applications may have automated tests (which we cannot confirm due to their closed source nature), they may be positive tests which *"make their program work rather than breaking the code"* as would be the case with negative tests. In our interviews, we questioned the participants on whether their application makes use of any kind of testing (**Q9**). Our results show that for some of these applications a simple mutation such as malforming the web API response caused a crash. Considering the *OZsale* application as an example, the interviewed mobile software architect claimed they do perform automated testing for some bad scenarios which may potentially happen, this very same application would remain loading indefinitely when faced with a malformed response.

## V. THREATS TO VALIDITY

**External validity.** Our study, despite including data for 48 applications, is composed solely of Android applications. Other mobile platforms which make use of web APIs such as iOS or Windows Phone should also be explored. Perhaps in some platforms it is more or less difficult to cause the whole application to crash. Also in some cases it is the same development team who develop for multiple platforms which use different programming languages and third party libraries for web API handling (e.g. HTTP libraries).

Similarly, our study can only be applied to mobile applications which make use of insecure HTTP to transport the web API requests and responses. This both limits the number of mobile applications which can be used as well as it may influence the outcome of the mutations. Mobile developers who intentionally chose to use HTTPS over HTTP are perhaps more conscious regarding the differences which make *web*

APIs different from the non-web counterparts. Indeed, without modifying the Android platform itself, nothing can be done to mitigate this threat.

**Reliability validity.** While our study intercepts and mutates web API responses and analyzes mobile applications' reactions to these mutations, we did not consider whether these mobile applications send failure data back to the respective software developers for further analysis. Such data, should it exist, may compliment and aid the debugging task.

## VI. RELATED WORK

**Maintenance of service-based systems.** Lewis and Smith were among the first to recognize that maintenance of service-based software systems is different from maintaining other types of software systems [21]. In particular, they highlight the importance of *impact analysis* for service providers as they have to consider a potentially unknown set of users.

Espinha et al. address this lack of knowledge regarding the user-base of services by tracking how different users use a service-based system in different ways [22].

Maleshkova et al. study the state of the practice on what concerns web API implementation and amongst the findings, discovered that the majority of the web APIs are actually underspecified [12].

**Evolution of APIs.** Li et al. [16] highlight the added challenges of web APIs versus statically linked APIs and provide a set of potential changes which web APIs may implement. The authors perform a similar study on how web service API evolution affects clients by analyzing what are common changes applied to web APIs and what these changes represent to client code in terms of API change patterns. Ultimately the authors propose the creation of a tool for automated migration.

Dig and Johnson try to understand the nature of changes to APIs [6]. From the five case studies that they analyzed in detail, they found that over 80% of the API-breaking changes can be classified as being refactorings.

McDonnell et al. through a study on API stability and adoption in the Android ecosystem have found that, despite the added benefits of newer versions of APIs, developers tend to be slow in adopting the newer versions [23], thus further highlighting the awareness required when web API changes are inevitable.

An interesting non-peer reviewed work in this field is a survey [24] conducted on the pains of web API integration which presents many complaints from web API client developers.

Daigneau focuses on the brittleness of web APIs in his book on service design patterns [25]. He proposes the *Single Message Argument* pattern, which suggests to refrain from creating signatures with long parameter lists. Daigneau further states that long parameter lists *"[...] signal the underlying framework to impose a strict ordering of parameters which, in turn, increases client-service coupling and makes it more difficult to evolve the client and service at different rates."*

## VII. CONCLUSION

In this paper we perform a study on the impact that changes to web API behavior can have on mobile applications. Our contributions are:

- An approach using mutation analysis for simulating unexpected responses from web APIs.
- A study on how 48 high profile mobile applications react to a set of predefined mutations in web API responses.
- Insight on caching and versioning approaches of some of the web APIs under study.
- An interview with three developers of some of the studied mobile applications.

Referring back to our research questions proposed in the introduction, we set out to find how robust mobile applications are when facing unexpected responses from web APIs. The first question we answer is **[RQ1.1]** on *"how can we simulate unexpected responses from web APIs"*. The mutation analysis presents a structured approach to simulate web APIs afflicted either by failure or by changes caused by software evolution.

Using mutation analysis we are then able to address **[RQ1]** which asks *"how robust are mobile apps when the web APIs being used return unexpected responses?"*. Our results present a mixed answer to this question. Indeed, most of the mobile applications studied are fairly robust to mutations in the web API response as seen by only 25% of the applications studied crashing through one of the mutations. Nonetheless, some of the mobile applications are not as resilient, as some applications crashed or silently failed upon facing changes to the web API. This behavior should be made more informative and user-friendly, which can be achieved through better understanding potential changes to web APIs.

Also **[RQ2]** which asks *"have web API client developers developed resilience against changes in the web API or failure of the web API?"* is answered with mixed results. Some of the applications studied make use of state of the art approaches (e.g. the HATEOAS versioning) to ensure a smooth evolution of their web API client, where others do not use versioning altogether (which as reported in previous work [8] may cause long-term pains) and allow the application to crash. The need for this resilience exists also outside of the source code. One of the interviewed developers raised concerns with inter-team communication, highlighting the need for clear and concise documentation from web API providers to client developers.

Our main research question asks *"how well-prepared are Android mobile applications with regard to changes in response messages from the web API"*. We conclude that while the majority of the studied applications are capable of dealing with such changes without major issues, some applications still use web APIs as if their behavior can be expected to never change, which as we have seen does not always happen.

**Future work.** We aim to extend our investigation to both paid mobile applications as well as iOS applications. The underlying platform (Android) may provide more or less support for web API integration, which we would like to investigate in the future.

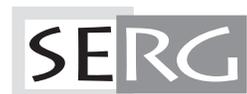